\def\half{\frac{1}{2}}
\def\mb#1{\mbox{\boldmath{$#1$}}}
\def\dbox#1{\hbox{\vrule  
                        \vbox{\hrule \vskip #1
                             \hbox{\hskip #1
                                 \vbox{\hsize=#1}%
                              \hskip #1}%
                         \vskip #1 \hrule}%
                      \vrule}}
\def\qed{\hfill \dbox{0.05true in}}  
\documentclass[showkeys,showpacs]{elsarta}
\usepackage{graphicx}
\begin{document}

\begin{frontmatter}

\markboth{R.~L.~Hall \& W.~Lucha}{Klein--Gordon lower bound to the semirelativistic ground-state energy}

\title{Klein--Gordon lower bound to the semirelativistic ground-state energy}

\author{Richard~L.~Hall}
\address{Department of Mathematics and Statistics, Concordia
University,\\ 1455 de Maisonneuve Boulevard West, Montr\'eal,
Qu\'ebec, Canada H3G 1M8.}
\ead{rhall@mathstat.concordia.ca}

\author{Wolfgang~Lucha}
\address{Institute for High Energy Physics, Austrian Academy
of Sciences,\\ Nikolsdorfergasse 18, A-1050 Vienna, Austria.}
\ead{wolfgang.lucha@oeaw.ac.at}

\begin{abstract}For the class of attractive potentials $V(r) \le 0$ which vanish at infinity,
we prove that the ground-state energy $E$ of the semirelativistic Hamiltonian
 $H = \sqrt{m^2 + p^2} + V(r)$ is bounded below by the ground-state
energy $e$ of the corresponding Klein--Gordon problem 
 $(p^2 + m^2)\phi = (V(r) -e)^2\phi.$ Detailed results are presented for the exponential and Woods--Saxon potentials.
\end{abstract}

\end{frontmatter}
\maketitle
\section{Introduction}
We study the ground states of a bound one-particle system in two different theories: the spinless-Salpeter equation \cite{BSE,SE,Lieb96,LS99,LS04}, and the relativistic Klein--Gordon equation \cite{Greiner}.  The spinless--Salpeter equation constitutes a well-defined
approximation to the Bethe--Salpeter formalism for the description
of bound states within (relativistic) quantum field theory. It may
be deduced from the Bethe--Salpeter equation by assuming that (a)
the interactions between the involved bound-state constituents are
instantaneous --- which leads to the so-called static limit ---
and (b) all bound-state constituents propagate like free
particles, and by neglecting both negative-energy contributions
and the spin degrees of freedom of the bound-state constituents.
On an equal footing, the spinless-Salpeter equation may be
regarded as the simplest conceivable generalization of the
nonrelativistic Schr\"odinger equation towards the incorporation
of relativistic dynamics.

In units with $\hbar = c = 1,$ if $\Phi(\mb{r}) \in L^2(R^3) $ is a bound-state wave function, and the operator $\mb{p}^2 = -\Delta,$ then for states corresponding to $\ell = 0 $ we may define the corresponding radial function $\phi(r) \in L^2(R^{+})$, where $\phi(r) = \sqrt{4\pi}\, r\, \Phi(\mb{r})$, and $r = |\mb{r}|.$ Since 
$$\Delta\Phi(\mb{r}) = \frac{\phi''(r)}{\sqrt{4\pi}\,r},$$
we now write the kinetic-energy operator as $p^2$, where $p^2\phi(r) = -\phi''(r). $  In this  notation,  the radial eigenequations corresponding to an attractive central vector potential $V(r)$, and zero scalar potential $S(r) = 0$, may be written respectively as
\begin{equation}\label{Eqsal}
H\psi = E\psi,\quad H = \sqrt{p^2 + m^2} + V(r),
\end{equation}
\begin{equation}\label{Eqkg}
h(e)\phi = (e^2-m^2)\phi,\quad h(e) = p^2 + 2eV(r) - V^2(r).
\end{equation}
The normalization condition for the radial functions is explicitly $\int_0^{\infty}\psi^2(r)\,dr =\int_0^{\infty}\phi^2(r)\,dr = 1.$  When the Klein--Gordon radial equation~(\ref{Eqkg}) is written in this way, $h(e)$ becomes a Schr\"odinger operator depending on a real parameter $e$: for suitable $V(r)$ and values of $e$, $h(e)$ has a discrete lowest eigenvalue, which we write as $F(e).$  The parameter $e$ becomes a Klein--Gordon eigenvalue when it satisfies $F(e) = e^2 - m^2:$ we may think of this as an intersection point of two graphs, that of $F(e)$ and the graph of $e^2-m^2$; thus there is a Klein--Gordon eigenvalue if and only if these two graphs intersect; the lowest Klein--Gordon eigenvalue is the smallest value of $e$ over all such intersection points.  We shall consider some examples below.  We emphasize that in our later discussions, $F(e)$ always represents the lowest eigenvalue of the Schr\"odinger operator $h(e)$.  
\medskip

The potentials we consider are of the form
\begin{equation}\label{Eqpot}
 V(r) = vf(r),
\end{equation}
where $f(r) \le 0$ is the potential shape, and $v > 0$ is a coupling parameter. We shall assume that the shape $f(r)$ is non decreasing, no more singular than Coulomb at $r=0,$ and vanishes as $r\rightarrow \infty.$ The semirelativistic and Klein--Gordon problems share some features and differ in others.  They both support discrete eigenvalues for the Coulomb potential $f(r) = -1/r$ only if the coupling is not too large: for example, with $m=1$ we require for the semirelativistic problem \cite{Herbst} $v < 2/\pi$ and for the Klein--Gordon equation $v < \half$; we note parenthetically that for the Dirac equation (in which $-v/r$ is the vector potential), we require $v < 1,$ corresponding to $Z < 137$ and $v = Z\alpha$.  For the exponential potential $V(r) = -ve^{-r},$ the two equations that we now study both require the coupling $v$ to be sufficiently large to give binding, but, whereas the semirelativistic equation has a discrete eigenvalue for arbitrarily large values of $v$, in the Klein--Gordon case (again with $m  = 1$), if $v$ is increased beyond about $v = 5.67,$ the ground state $\phi$  soon becomes `supercritical' as $e$ attempts to venture into the region $e < -m.$  In fact, for the Klein--Gordon problem with this class of attractive potentials, the allowed coupling is generally restricted by the requirement \cite{Greiner} that $-m < e < m.$ Thus, in comparing $E$ with the corresponding $e$ for the same potential and mass, we shall restrict the coupling $v$, above and below, so that both eigenvalues exist.  With these assumptions, we can state the principal result of this paper, namely that $E \ge e.$ When $e >0,$ this result is not difficult to establish; for the general case, including negative $e$, we shall need to prove that with $v$ fixed, $e(m)$ increases with $m.$

We note that for unbounded potentials, quite outside the considered class, such as the harmonic oscillator $f(r) = r^2,$ the semirelativistic problem has bound states \cite{HLS04}, but the corresponding Klein--Gordon problem has no discrete spectrum at all.  Such problems are therefore not compatible with the main purpose of this paper and will not be discussed further here.

\begin{figure}[htbp]\centering\includegraphics[width=12cm]{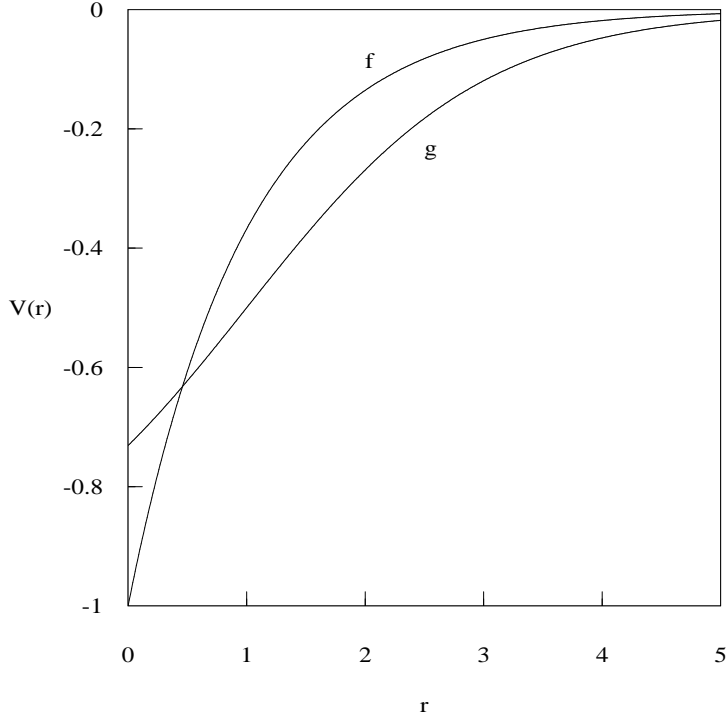}
\caption{The exponential and Woods--Saxon pair-potential shapes $f(r) = -e^{-r}$ and $g(r) = -1/(1+e^{(r-1)})$ in dimensionless units.}\label{Fig:fg}\end{figure}

In the next section we state more precisely what our assumptions are and we set out our main results in two theorems and a convexity lemma. We provide some illustrations in that section in terms of the exponential potential $f(r) = -e^{-r}.$ In section~3 we look at the Woods--Saxon nuclear potential $V(r) = -v/[1+e^{(r-a)/b}]$ \cite{WS} and provide upper and lower energy bounds to the semirelativistic energy $E$, as functions of the coupling $v.$ In Fig.~\ref{Fig:fg} we illustrate these two potential shapes as examples of the class of potentials we are studying.

\section{The lower energy bound}

We shall find it helpful to define the value of the parameter $e$ that gives rise to a zero eigenvalue for $h(e)$ of Eq.~(\ref{Eqkg}) to be $e_0$: thus $F(e_0) = 0.$
We must now characterize the $F(e)$ curves more generally.  We shall first show in general that $F(e)$ is negative, decreasing, and concave.  In order to do this it is simplest to fix the parameter $v$ suitably and think of the solution to $e^2-m^2 = F(e)$ as the function $e(m).$  Since the potentials $V(r)$ we are considering are negative and vanish at infinity, whenever the eigenvalue $F(e)$ of $h(e)$ exists, it must be negative.  Since the Klein--Gordon energies $e$ are solutions of $F(e) = e^2 -m^2,$ if follows that $-m < e < m$, a result discussed by Greiner \cite{Greiner}. We write the normalized Klein--Gordon eigenfunction as $\phi(r,e)$, which, for real potentials, we may without loss of generality assume to be real. We indicate partial derivatives with respect to the parameter $e$ by a subscript, thus $\partial\phi/\partial e = \phi_e.$  By differentiating $(\phi,\,\phi) = 1$ with respect to $e$ we obtain the orthogonality relation $(\phi,\,\phi_e) = 0.$ We now differentiate the equation $F(e) = e^2-m^2 = (\phi, h\phi)$ with respect to $m$ to obtain
\begin{eqnarray}\label{Eqdifh}
\nonumber F'(e(m))e'(m)&=& 2e(m)e'(m)-2m \\
&=& \left[(\phi_e,\,h\phi) + (\phi,\,2V(r)\phi) + (\phi,h\,\phi_e)\right]e'(m).
\end{eqnarray}
By using the self-adjointness of $h(e)$ and the orthogonality relation, we find from ~Eq.~(\ref{Eqdifh}) that $F'(e) = (\phi, \,2V(r)\phi) \leq 0$ (since $V(r) \leq 0$), and moreover
\begin{equation}\label{Eqeprime}
e'(m) = \frac{m}{e(m) -\half F'(e(m))} = \frac{m}{\delta(e)}.
\end{equation}
It follows immediately from Eq.~(\ref{Eqeprime}) that if $e(m) \ge 0 ,$ then $e'(m) \ge 0.$  We have studied an example of this earlier \cite{HL06,HL08}, namely the Coulomb (or gravity) case $V(r) = -v/r$, with $v$ not too large: for this problem, the coupling $v$ can never be large enough to generate negative Klein--Gordon eigenvalues $e$ and consequently we predict that $e'(m) > 0$ (as indeed we know from the exact solution), and therefore we know that $E > e.$ In this earlier work we could solve the Kratzer--Schr\"odinger problem \cite{Kratzer, Schroedinger} exactly for $F(e)$ and, in turn, $e$ could be found by solving $F(e) = e^2-m^2$ algebraically. Thus a more general theory was unnecessary for that particular problem. But, to return to the present task, we still need to prove generally that $e'(m) > 0$ even when $e$ is negative.  We first establish the concavity of $F(e)$ by proving the lemma

\medskip
\noindent{\bf Lemma~1:} {\it  $F(e)$ is concave.}

\medskip
\noindent{\bf Proof:}~~We suppose that $\phi$ and $\phi_1$ are respectively the normalized ground states of $h(e)= p^2 + 2eV -V^2$ and $h(e_1) = p^2 + 2e_1V -V^2$. 
We therefore have
\begin{eqnarray}\label{Eqconcave}
\nonumber F(e) \le (\phi_1, h(e) \phi_1) &=& (\phi_1,h(e_1)\phi_1) +(e -e_1)(\phi_1,2V\phi_1)\\
\nonumber &=&F(e_1) +(e -e_1)F'(e_1).
\end{eqnarray}
Thus $F(e)$ lies beneath its tangents and is concave.\qed\medskip

A function that is important for the theory is $\delta(e) = e -\half F'(e).$ Since $F(e)$ is concave, we may assume $\delta'(e) = 1 -\half F''(e)  > 1.$ We now state and prove the first theorem.

\medskip
\noindent{\bf Theorem~1:} {\it Let $E$ be the lowest eigenvalue of the spinless--Salpeter equation (\ref{Eqsal}) and $e$ be the lowest eigenvalue of the Klein--Gordon equation (\ref{Eqkg}), then if $e \ge 0$, or $e < 0$ and $\delta(e) > 0$, it follows that $E \geq e$.}

\begin{figure}[htbp]\centering\includegraphics[width=12cm]{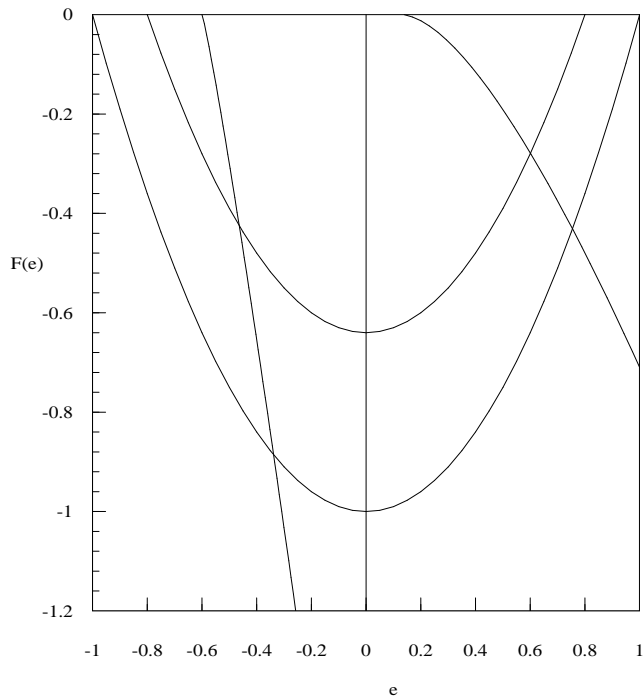}
\caption{The exponential pair potential $V(r) = -v\,e^{-r}.$ The curves are the Schr\"odinger energies $F(e)$ of the operator $h(e) = p^2 +2eV-V^2$ for $v = 2.5 $ and $v = 4.5$, along with the curves $g(e) = e^2 - m^2$ for $ m = 0.8$ and $m = 1$, in units with $\hbar = c = 1.$  The intersection points yield the Klein--Gordon ground-state energies $e = e(m,v).$ For these solutions to exist, the $F(e)$ curves must be steeper than the $g(e)$ curves: thus we have $F'(e) < g'(e) = 2e$ at the intersection points, even when $e$ is negative; this is a sufficient condition for the Klein--Gordon energy $e$ to be a lower bound to the corresponding semirelativistic energy $E>e$.}\label{Fig:F(e)exp}\end{figure}

\begin{figure}[htbp]\centering\includegraphics[width=12cm]{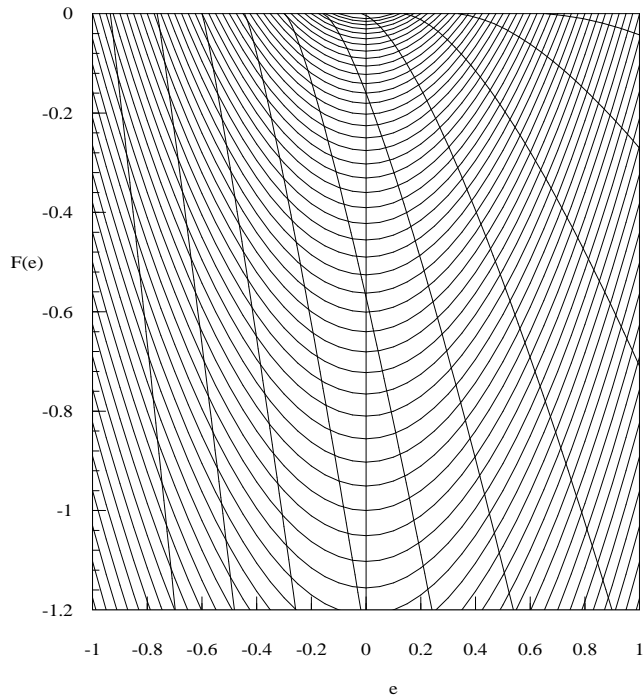}
\caption{The curves of Schr\"odinger energies $F(e)$ for $v = 1\,(0.5)\,5.5$ and $g(e) = e^2 - m^2$ for $ m = 0.1\,(0.025)\,2.1$ for the exponential pair potential $V(r) = -v\,e^{-r}$, in units with $\hbar = c = 1.$  The intersection points yield the Klein--Gordon ground-state energies $e = e(m,v).$ For these solutions to exist, the $F(e)$ curves must be steeper than the $g(e)$ curves: thus we have $F'(e) < g'(e) = 2e$ at the intersection points, even when $e$ is negative; this is a sufficient condition for the Klein--Gordon energy $e$ to be a lower bound to the corresponding semirelativistic energy $E>e$.}
\label{Fig:F(e)expn}\end{figure}

\medskip
\noindent{\bf Proof:}~~With the potential parameters fixed and $m$ sufficently large, we can see from the Klein--Gordon equation (\ref{Eqkg}) that we certainly obtain a solution if the Schr\"odinger operator $h(e)$ has eigenvalues $F(e)$ that are negative.  For the present consideration we re-write the Klein--Gordon equation as
\begin{equation}\label{Eqkg2}
h(e)\phi = \left(p^2 + 2eV(r) -V^2(r)\right)\phi = F(e)\phi = (e^2 -m^2)\phi.
\end{equation}
A Klein--Gordon eigenvalue $e$ is a parameter value in $h(e)$ such that the graph of $F(e)$ intersects the graph of $e^2-m^2$. For this to happen, $m$ must be sufficiently large, and such a solution lies in the interval $e \in(-m,\,m).$ Meanwhile, with the same set of (compatible) potential parameters, if we write the semirelativistic equation (\ref{Eqsal}) in the form $\sqrt{p^2+m^2}\,\psi = (E-V(r))\psi$, and then square the equal vectors, we obtain 
\begin{equation}
E^2  -  m^2 = \left(\psi,\ \left(p^2 + 2EV(r) - V^2(r)\right)\psi\right) \geq F(E),\quad (\psi,\ \psi) =1.
\end{equation}
The inequality arises from the variational principle applied to the Schr\"odinger operator $h(E)$ in which the parameter $e$ has been set to the value $E.$
 Thus we have $E^2 - m^2 \geq F(E).$  A slack variable $M\ge m$ can be introduced to convert this last inequality into the equality
$E^2 -M^2 = F(E)$.  We may now summarize the relation between the Klein--Gordon energy $e$ and the semirelativistic energy $E$ corresponding to
 a  compatible attractive potential $V(r)$ by the equations:
\begin{equation}\label{Eqkgcom}
e^2-m^2 = F(e), \quad E^2 - M^2 = F(E),\quad M\ge m.
\end{equation}
Our main result $E \geq e$ will be established if we can find conditions sufficient to show that the solution $e(m)$ of the Klein--Gordon problem $e^2-m^2 = F(e)$ 
increases with $m$.  We first obtain an expression for $e'(m)$ and then consider the two cases $e \ge 0$ and $e < 0.$
In order to fix ideas, we first consider an example.  In Fig.~\ref{Fig:F(e)exp} we have plotted the spectral curves $F(e)$ for the potential $V(r) = -v\,\exp(-r)$ and $v = 2.5$ and $v = 4.5$ (the left and right falling $F(e)$ curves). We have also plotted the U-shaped curves representing $g(e) = e^2-m^2$ with $m= 1$ and $m = 0.8.$ The Klein--Gordon energy is the $e$ value at the intersection of an $F(e)$ and a $g(e)$ graph.  When we follow each $F(e)$ curve down in Fig.~\ref{Fig:F(e)exp}, it is clear that the two solutions in the example both satisfy $e(1) > e(0.8).$ In Fig.~\ref{Fig:F(e)expn} we show a larger family of curves for the exponential potential; in Fig.~\ref{Fig:F(e)woodssaxon} we exhibit corresponding curves for the Woods--Saxon potential (which we discuss later, in section~3). The situation is similar for the whole class of potentials which we study in this paper.

We now consider the case $e < 0.$ It follows from the concavity of $F(e)$ that $\delta'(e) = 1 - \half F''(e) > 1.$ Thus, if $\delta(e)>0,$ then, as $e$ increases, $\delta(e)$ remains  positive.  Since $\delta(e) > 0,$ we know from Eq.~(\ref{Eqeprime}) that $e'(m) > 0$. Meanwhile, $\delta'(e) > 1.$ These facts tell us that, as $e$ increases, $\delta(e)$ remains positive, and so does $e'(m)$. 
If $M > m,$ then $E = e(M) > e(m).$ This proves Theorem~1.
\qed\medskip

Theorem~1 supposes that the Klein--Gordon solution $e < 0$ satisfies $\delta(e) > 0.$  We now explore some sufficient conditions for this to happen, based on a certain critical value $e_0$ of $e$. We suppose that, for a given potential $V$, there is a value $e_0$ of $e$ such that $F(e_0) = 0.$ If $e_0 \ge 0,$ then solutions of $g(e) = e^2 - m^2 = F(e)$ satisfy $e > 0.$ We therefore do not discuss this possibility further, but rather assume that $e_0 < 0.$  We now consider two cases: (i) $e_0\le -m$ and (ii) $e_0 > -m.$  In case (i), if $\delta(e_0) \le 0,$ then there could be two Klein--Gordon solutions, $\{e_1,\,e_2\}$ with $e_1 < e_2$: in this event, $\delta(e_2) > 0,$ and we know by Theorem~1 that $E > e_2.$ Meanwhile, if $\delta(e_0) >0,$ there are no solutions because the graphs of $F(e)$ and $g(e)$ have opposite convexities and never meet.  This leaves us with case (ii), which allows us to prove:

\medskip
\noindent{\bf Theorem~2:} {\it Let $E$ be the lowest eigenvalue of the spinless--Salpeter equation (\ref{Eqsal}) and $e$ be the lowest eigenvalue of the Klein--Gordon equation (\ref{Eqkg}), then, if $e \ge 0$, or $0 > e_0 > -m,$ then $E \geq e$.}

\medskip
\noindent{\bf Proof:}~~Since the case $e \ge 0$ was proved generally for Theorem~1, we suppose that $e < 0.$ Since $-m < e_0 < 0,$ the $F(e)$ curve crosses the $g(e)$ curve (as $e$ increases) from above. Consequently, at the intersection we must have $F'(e) < g'(e)$, that is to say, $F'(e) < 2e,$ or $\delta(e) > 0.$  Therefore, we know from Theorem~1 that $E = e(M) \ge e(m).$ 
\qed\medskip


\section{The Woods--Saxon potential}
\begin{figure}[htbp]\centering\includegraphics[width=12cm]{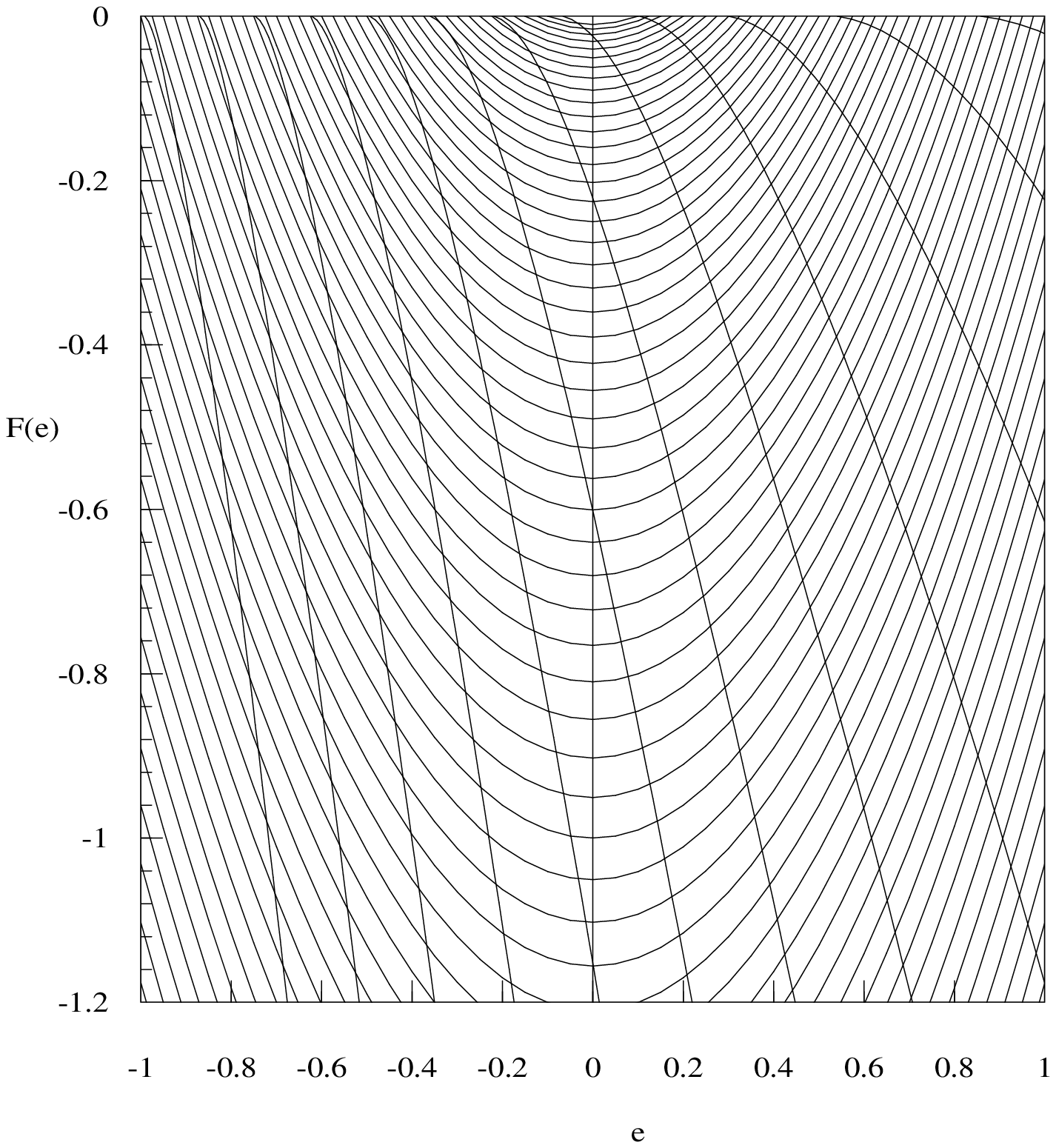}
\caption{The curves of Schr\"odinger energies $F(e)$ for $v = 1\,(0.25)\,3.75$ and and $g(e) = e^2 - m^2$ for $ m = 0.1\,(0.025)\,2.1$ for the Woods--Saxon pair potential $V(r) = -v\,/[1+ e^{5(r-1)}]$, in units with $\hbar = c = 1.$ The intersection points yield the Klein--Gordon ground-state energies $e = e(m,v).$ For these solutions to exist, the $F(e)$ curves must be steeper than the $g(e)$ curves: thus we have $F'(e) < g'(e) = 2e$ at the intersection points, even when $e$ is negative; this is a sufficient condition for the Klein--Gordon energy $e$ to be a lower bound to the corresponding semirelativistic energy $E>e$.}
\label{Fig:F(e)woodssaxon}\end{figure}

We now look at the Woods--Saxon nuclear potential $V(r) = -v/[1+e^{-(r-a)/b}]$.  In Fig.~\ref{Fig:F(e)woodssaxon} we plot for the Woods--Saxon potential, with parameters $\{a = 1,\, b = 1/5\}$, the $F(e)$ and $g(e)$ curves whose intersection points provide Klein--Gordon energies $e.$ It is clear from these graphs that the condition $F'(e) < g'(e) = 2e$, sufficient for us to know that $e$ is a lower bound to $E$, is indeed satisfied in this case. Thus we have a lower energy bound $e \le E.$

\begin{figure}[htbp]\centering\includegraphics[width=12cm]{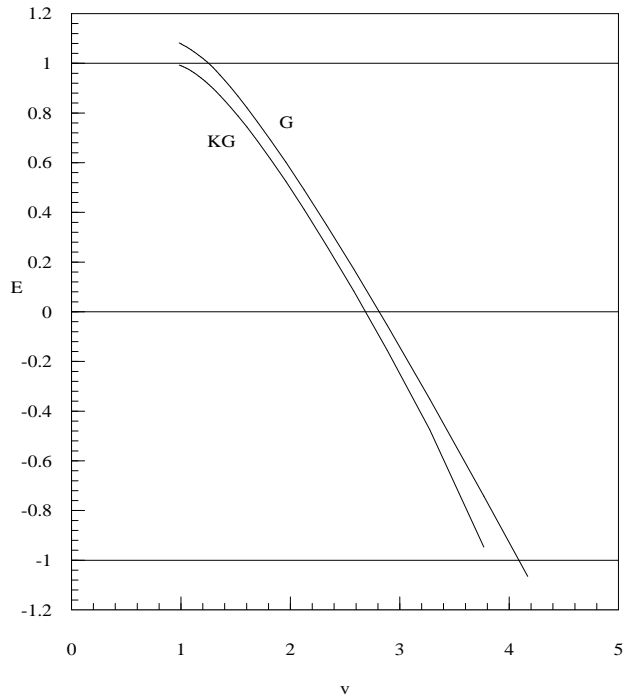}
\caption{Energy bounds $e(v)<E(v) < E_g(v)$ for the spinless--Salpeter problem with the Woods--Saxon potential $V(r) = -v/[1+e^{5(r-1)}]$, in units with $\hbar = c = 1.$ The lower bound is given by the corresponding Klein--Gordon energy (KG), and an upper bound to the spinless--Salpeter energy (G) is provided by  use of a scale-optimized Gaussian wave function applied to the spinless--Salpeter problem itself.}\label{Fig:energybounds}\end{figure}

If we follow the upper bound treatment of \cite{HL06} we have for
\begin{equation}\label{EqHam}
H = \sqrt{p^2 + m^2} - \frac{v}{1+e^{(r-a)/b}}
\end{equation}
and a Gaussian trial function given by
\begin{equation}\label{EqWavefn}
\phi(r) =\left(\frac{1}{s^2\pi}\right)^{\frac{3}{4}} e^{-\half(r/s)^2}
\end{equation}
that, for each fixed parameter set $\{m,a,b,v\}$, the Gaussian upper
energy bound $E_g \ge E$ is given by
\begin{equation}\label{EqEg}
E_g =\int\limits_0^{\infty}\rho(t)\left[\frac{(m^2s^2 + t^2)^{\half}}{s} - \frac{v}{1+e^{(ts-a)/b}}\right]dt=J_1(s) - vJ_2(s),
\end{equation}
where the probability density $\rho(t)$ on $[0, \infty)$ is defined by
\begin{equation}\label{Eqdensity}
\rho(t) = \frac{4}{\pi^{\half}}\,t^2 e^{-t^2}.
\end{equation}
If we keep the parameters $\{m,a,b\}$ fixed and optimize over the scale $s$ for each value of the coupling $v$, we obtain the parametric equations for the best Gaussian upper-bound curve $\{v,\,E_g(v)\}$ in the form
\begin{equation}\label{Eqpar}
E_g = J_1(s) - vJ_2(s), \quad v = J_3(s)/J_4(s),\, s > 0,
\end{equation}
where
\begin{equation}\label{EqJ3}
J_3(s) = \frac{1}{s^2}\int\limits_0^{\infty}\rho(t) \frac{t^2}{(m^2s^2 + t^2)^{\half}}dt
\end{equation}
and
\begin{equation}\label{EqJ4}
J_4(s) = \frac{1}{b}\int\limits_0^{\infty}\rho(t)\frac{te^{(st-a)/b}}{[1+e^{(st-a)/b}]^2}\,dt.
\end{equation}
For the case $\{ m = a = 1,\,b = 1/5\}$ we have plotted in Fig.~\ref{Fig:energybounds} the energy bounds $\{e(v),\,E_g(v)\}$ as functions of the coupling $v.$


\section{Conclusion}
We have shown that the type of argument  used to establish a lower energy bound for the semirelativistic eigenvalue problem with an attractive Coulomb (or gravitational) potential \cite{HL08}, may also be  used for a wide class of attractive potentials that are negative and increase monotonically to zero at infinity.  The analysis of the Klein--Gordon spectral problem  in terms of the Schr\"odinger sub-problem may itself be of interest to researchers in the field. It is also hoped that the main result will be useful in providing lower energy bounds for semirelativistic $N$-particle systems bound together by pair potentials of the type discussed in this paper.

\section*{Acknowledgements}

One of us (RLH) gratefully acknowledges both partial financial support
of this research under Grant No.\ GP3438 from the Natural Sciences
and Engineering Research Council of Canada, and the hospitality of
the Institute for High Energy Physics of the Austrian Academy of
Sciences, Vienna, where part of the work was done.

\end{document}